\documentclass[prc,aps,preprint,showpacs,groupedaddress, superscriptaddress]{revtex4}

\usepackage{latexsym,graphicx,epsfig,color,xspace}
\usepackage{amssymb,amsmath,amsxtra,amsfonts}
\usepackage{multirow}

\usepackage{bm}

\usepackage{titlesec}

\titleformat{\section}{\large\bfseries}{\thesection}{1em}{}

\newcommand{\bea}{\begin{eqnarray}}
\newcommand{\ena}{\end{eqnarray}}
\newcommand{\be}{\begin{equation}}
\newcommand{\en}{\end{equation}}

\newcommand{\ed}{\end{document}}

\begin{document}

\title{Investigation of low-lying resonances in breakup of halo nuclei within the time-dependent approach}

\author{D.S. Valiolda}
\affiliation {Bogoliubov Laboratory of Theoretical Physics, Joint Institute for Nuclear Research, Dubna, Moscow Region 141980, Russian Federation}
\affiliation{al-Farabi Kazakh National University, Almaty 050040, Kazakhstan}
\affiliation{Institute of Nuclear Physics, Almaty 050032, Kazakhstan}

\author{D.M. Janseitov}
\email{janseit.daniar@gmail.com}
\affiliation {Bogoliubov Laboratory of Theoretical Physics, Joint Institute for Nuclear Research, Dubna, Moscow Region 141980, Russian Federation}
\affiliation{al-Farabi Kazakh National University, Almaty 050040, Kazakhstan}
\affiliation{Institute of Nuclear Physics, Almaty 050032, Kazakhstan}

\author{V.S. Melezhik}
\email{melezhik@theor.jinr.ru}
\affiliation {Bogoliubov Laboratory of Theoretical Physics, Joint Institute for Nuclear Research, Dubna, Moscow Region 141980, Russian Federation}
\affiliation {Dubna State University, Moscow Region 141982, Russian Federation}

\begin{abstract}

We investigate the Coulomb breakup of $^{11}$Be halo nuclei on a heavy target from intermediate (70 MeV/nucleon) to low energies (5 MeV/nucleon) within the non-perturbative  time-dependent approach. The convergence of the computational scheme is demonstrated in this energy range including  \textit{n}+$^{10}$Be low-lying resonances in different partial and spin states. We have found a considerable contribution of the $5/2^{+}$ resonance ($E_{r}$= 1.23 MeV) to the breakup cross section at 30 MeV/nucleon and lower, while at higher energies, the resonant states $3/2^{-}$ and $3/2^{+}$ (with $E_{r}$= 2.78 and 3.3 MeV) make most visible contributions. The obtained results are in good agreement with experimental data available at 69 and 72 MeV/nucleon. Comparison with the existing theoretical calculations of other authors for 20 and 30 MeV/nucleon is also made. The developed computational scheme opens new possibilities in the investigation of the Coulomb, as well as nuclear, breakup of other halo nuclei on heavy and light targets.

\end{abstract}


\maketitle

\section{Introduction}
\label{sec:intro}

Exotic structures and phenomena in the proximity of the proton and neutron driplines were discovered in the 1980s with radioactive beams. A striking example is the halo nuclei. Due to their very short lifetime, halo nuclei are studied through indirect techniques such as breakup reactions in which the halo nucleus dissociates from the core through interaction with the target. A number of breakup measurements were made even when the radioactive beam intensity was rather low. The Coulomb breakup is one of the main tools here~\cite{JAH:(2004),JPG:22(1996)}.

Halo nuclei are usually described within the few-body model: a core which contains most of the nucleons plus one or two valence nucleons. A number of computational approaches were developed to describe the breakup of two-body projectiles, i.e., one-nucleon halo nuclei: perturbation expansion~\cite{PRC50:2104(1994), NPA:600(1996)}, adiabatic approximation~\cite{PRC:57(1998)}, eikonal model~\cite{SLYV:(2003)}, coupled-channels with a discretized continuum (CDCC)~\cite{TPS:89(1986), PRC:63(2001)}, dynamical eikonal approximation (DEA)~\cite{PRL:95(2005)} and numerical integration of the three-dimensional time-dependent Schr\"odinger equation (TDSE)~\cite{PRC:59(1999), PRC:68(2003), PRC50:1276(1994), NPA:581(1995), PRC:53(1996)}. In some of these works both the Coulomb and nuclear breakup mechanisms were considered.

The methods mentioned above are mainly applied to high and intermediate beam energies in the breakup reaction of $^{11}$Be on a heavy target.
For intermediate beam energies (near 70 MeV/nucleon) there are rather accurate experimental data~\cite{PLB:331(1994), PRC70:054(2004)} and theoretical analyses with different approaches~\cite{PRL:95(2005), PRC:59(1999), PRC:68(2003), PRC50:1276(1994), NPA:581(1995), PRC:53(1996), PRC70:044(2004), PRC:89(2014), JPG:42(2015)}. However, for lower energies only a few theoretical works have been performed so far~\cite{PRL:95(2005), PRC:65(2002), PRC:73(2006)}. For instance at~\cite{PRC:65(2002)} the breakup reaction $^{11}$Be+$^{208}$Pb $\rightarrow ^{10}$Be+n+$^{208}$Pb for energies of 5, 10 and 30 MeV/nucleon was investigated within the distorted-wave Born approximation (DWBA). Nevertheless, the applicability of this approach to low energies is questionable and must be justified~\cite{PRC:65(2002)}. An attempt was also made to calculate the breakup of $^{11}$Be on the Pb target at a beam energy of 20 MeV/nucleon with the dynamical eikonal approximation~\cite{PRL:95(2005), PRC:73(2006)}.
However, this approach is a high-energy reaction model~\cite{PRL:95(2005)} and its adequate extension to low-energy (below 40 MeV/nucleon) has not yet been realised~\cite{PRC:98(2018)}.
Thus, although the region around 10 MeV/nucleon is of great interest, since this is the energy range of HIE-ISOLD at CERN and the future ReA12 at MSU, it has hardly been investigated so far.

In the present work, we fill this gap by generalizing to this area the theoretical model developed in~\cite{PRC:59(1999), PRC:68(2003), PRC:64(2001)} and successfully applied to the breakup of halo nuclei $^{11}$Be~\cite{PRC:59(1999), PRC:68(2003)}, $^{15}$C~\cite{PRC:68(2003)}, and $^{17}$F~\cite{PRC:64(2001)} at higher beam energies. In this model, the time-dependent Schr\"odinger equation for a halo-nucleus is integrated with a non-perturbative algorithm on a three-dimensional spatial mesh. The use of the discrete-variable representation in 2D angular space and high-order finite differences for the radial part of the wave function allows avoiding the multipole expansion of the time-dependent Coulomb interaction between the projectile and the target~\cite{PRC:59(1999), VSM:(1998), PLA:230(1997)}.
Another  attractive feature of the method is its flexibility to the choice of the interactions between the halo-nucleus, the core and the target~\cite{PRC:68(2003)} and  in the definition of the projectile trajectory which can be classically treated  simultaneously with the Schr\"odinger equation for the weakly-bound halo-nucleus of the projectile~\cite{PRA:69(2004)}.

We extend the theoretical model to the low-energy region and investigate its convergence and stability here.
With this approach we perform calculations of the breakup cross section of $^{11}$Be  on a heavy target ($^{208}$Pb) at 5 - 30 MeV/nucleon taking into account Coulomb and nuclear interactions between the projectile and the target. We apply the classical consideration (classical trajectories) for a heavy and fast $^{11}$Be projectile and a quantum approach for a light and slow neutron, where the relation $p_{Be}=(3800-1100)$ MeV/c $>>$ $p_{n}\sim 40$ MeV/c for their momentums is satisfied. We consider the region $E_{Be} =(770-55)$ MeV $>>$ $E_{n}\sim 1 $ MeV for the projectile and the halo neutron energies during the breakup collisions, which further justifies the application of the classical description for the projectile.  We also analyse in the frame of this model the influence of the $^{11}$Be resonant states $5/2^{+}$, $3/2^{-}$ and $3/2^{+}$~\cite{PRC70:064(2004), PAN:77(2014), NNDC} on the breakup processes. This analysis demonstrates the possibility of studying low-lying resonances in halo nuclei using their breakup reactions. The method can potentially be useful for interpretation of low-energy breakup experiments on different targets in studying the halo structure of nuclei.

In Sec. II, the modeling of the physical problem is described. Also, the computational scheme and the conditions of the calculation are presented. The obtained results are discussed in Sec. III. The last section is devoted to concluding remarks.

\section{THEORETICAL FRAMEWORK}
\subsection{Model for describing neutron dynamics during collision}

The halo neutron is treated as a structureless particle weakly bound by the potential \textit{V(r)} to the $^{10}$Be core nucleus, where \textit{r} is the relative variable between the  neutron and the core. The dynamics of the halo neutron relative the $^{10}$Be core in the breakup reaction $^{11}$Be+$^{208}$Pb $\rightarrow ^{10}$Be+n+$^{208}$Pb  is described by the time-dependent Schr\"odinger equation

\begin{equation}
i\hslash\frac{\partial}{\partial\textit{t}}\Psi(\textbf{r},t)=H(\textbf{r},t)\Psi(\textbf{r},t)=[H_{0}(r)+V_{C}(\textbf{r},t)]\Psi(\textbf{r},t)
\label{one}
\end{equation}

\noindent in the projectile rest frame.
In this expression

\begin{equation}
H_{0}(r)=-\frac{\hslash^{2}}{2\mu}\Delta_{r}+V(r)
\label{one}
\end{equation}
is the Hamiltonian describing a relative halo nucleon-core motion with reduced mass $ \mu=m_{n}m_{c}/M$, where $m_{n}, m_{c}$ and M=$m_{n}+m_{c}$ are the neutron, $^{10}$Be-core, and $^{11}$Be masses, respectively. The potential \textit{V(r)} represents the sum of the \textit{l}-dependent central potential \textit{$V_{l}(r)$} and the spin-orbit interaction \textit{$V_{l}^{s}(r)(\textbf{ls})$}. The interaction of the target nucleus with the projectile corresponds to the time-dependent Coulomb potential \textit{$V_{C}$}$(\textbf{r},t)$, which is defined as

\begin{equation}
V_{C}(\textbf{r},t)=\frac{Z_{c}Z_{t}e^{2}}{|m_{n}\textbf{r}/M+\textbf{R}(t)|}-\frac{Z_{c}Z_{t}e^{2}}{R(t)}\,,
\label{one}
\end{equation}
where $Z_{c}$ and $Z_{t}$ are charge numbers of the core and target, respectively, and $\textbf{R}(t) $ is the relative coordinate between the projectile and the target, which represents the straight-line trajectory $\textbf{R}(t)=\textbf{b}+\textbf{v}_{0}t$, where $\textbf{b}$ is the impact parameter orthogonal to the initial velocity of the projectile $\textbf{v}_{0}$. Here we follow the definition accepted in the works~\cite{PRC:59(1999), PRC50:1276(1994), NPA:581(1995)}.

To solve the time-dependent four-dimensional Schr\"odinger equation (1), we use the non-perturbative approach suggested in~\cite{VSM:(1998), PLA:230(1997)} and successfully applied for a number of different few-body problems in muon~\cite{HI:101(1996)} and atomic~\cite{PLA:230(1997), VSM:(1998), PRA:69(2004)} physics, as well as for the breakup of halo nuclei~\cite{PRC:59(1999), PRC:68(2003), PRC:64(2001)}. Following~\cite{PRC:59(1999)}, we seek for a solution $\Psi(\textbf{r},t)$ of the Shr\"odinger Eq. (1) in the spherical coordinates $(r,\Omega)\equiv(r,\theta,\phi)$ as an expansion

\begin{equation}
\Psi(\textbf{r},t)=\frac{1}{r}\sum_{s}\sum_{\nu\textit{j}}^{N}\varphi_{\nu}(\Omega)(\varphi^{-1})_{\nu\textit{j}}\psi_{j}^{s}(r,t)
\label{one}
\end{equation}
over the two-dimensional basis

\begin{equation}
\varphi_{\nu}(\Omega)=\sum_{\nu^{'}=\lbrace\l^{'}m^{'}\rbrace}C_{lm}^{l^{'}m^{'}}P_{l^{'}}^{m^{'}}(\theta)e^{im^{'}\phi}\,.
\label{one}
\end{equation}
In this basis $C_{lm}^{l^{'}m^{'}}=\delta_{ll^{'}}\delta_{mm^{'}}$ in general, and thus $\varphi_{\nu}(\Omega)$ coincides with an usual spherical harmonic with a few possible exceptions for high $\nu$, as explained below after Eq. (7). The symbol $\nu$  represents (\textit{l,m}), and the sum over $\nu$ is equivalent to the double sum
\begin{equation}
\sum_{\nu=1}^{N}=\sum_{l=0}^{\sqrt{N}-1}\sum_{m=-l}^{l}\,.
\label{one}
\end{equation}
The basis (5) is associated with a mesh. For the $\theta$ variable, the $N_{\theta}$=$\sqrt{N}$ mesh points $\theta_{j_{\theta}}$ are defined as zeros of the Legendre polynomial $P_{\sqrt{N}}(cos\theta_{j_{\theta}})$. For the $\phi$ variable, the $N_{\phi}$=$\sqrt{N}$ mesh points are chosen as $\phi_{j_{\phi}}=\pi(2j_{\phi}-1)/\sqrt{N}$. The total number \textit{N} (where $N=N_{\theta}\times\textit{N}_{\varphi}$) of grid points $\Omega_{j}=(\theta_{j_{\theta}},\phi_{j_{\phi}})$  is equal to the number of basis functions (5) in expansion (4)~\cite{PLA:230(1997), PRA:69(2004)}. This mesh is associated with the \textit{N} weights $\lambda_{j}$ that are products of the standard Gauss-Legendre weights over $\theta$ by $2\pi/\sqrt{N}$. The $(\varphi^{-1})_{\nu\textit{j}}$ are the elements of the \textit{N}x\textit{N} matrix $\varphi^{-1}$ inverse to the matrix with the elements $\varphi_{j\nu}=\varphi_{\nu}(\Omega_{j})$ defined at the grid points $\Omega_{j}$. The polynomials $\varphi_{\nu}(\Omega)$ are constructed from the Legendre polynomials $ P^{m}_{l}(\theta) $ and $e^{im\phi}$ (see Eq. (5)) so that they are orthogonal on the grid $\Omega_{j}$
\begin{equation}
\int\varphi_{\nu}^{*}\varphi_{\nu^{'}}(\Omega)d\Omega=\sum_{j}\lambda_{j}\varphi_{\nu\textit{j}}^{*}\varphi_{\nu^{'}j}=\delta_{\nu\nu^{'}}
\label{one}
\end{equation}
for all $\nu$ and $\nu^{'}\leq N$. For most $\nu$ and $\nu^{'}$, property (7) is automatically satisfied because the basis functions $\varphi_{\nu}(\Omega)$ are orthogonal and the Gauss quadrature is exact. For these $\nu$ the coefficients $C_{lm}^{l^{'}m^{'}}=\delta_{ll}\delta_{mm}$ in Eq.(5). However, in a few cases with the highest \textit{l} and \textit{m} values, some polynomials $\varphi_{lm}(\Omega)$ have to be specially made orthogonal in the sense of the Gauss quadrature. With this choice, the matrix $\lambda^{1/2}_{j}\varphi_{\nu\textit{j}}$ is orthogonal.

The radial components $\psi_{j}^{s}(r,t)$ in expansion (4) correspond to $r\psi(r,\Omega_{j},t)\vert\textit{s}>$ where $\vert\textit{s}>=(\pm\frac{1}{2})$ is a spin state of a nucleus and $r\psi(r,\Omega_{j},t)$ are the values of the spatial part of the nuclear wave function at the angular grid points $\Omega_{j}$. With respect to the unknown coefficients in expansion (4), the problem is reduced to the system of Schr\"odinger-type equations

\begin{equation}
i\hslash\frac{\partial}{\partial\textit{t}}\bm{\psi}(r,t)=[\hat{H}_{0}(r)+\hat{h}(r,t)]\bm{\psi}(r,t)
\label{one}
\end{equation}

\noindent where the 2$N$-component vector $\bm{\psi}(r,t)$ is introduced as $\bm{\psi}(r,t)=\lbrace\lambda^{1/2}_{j}\psi_{j}^{s}(r,t)\rbrace$.

In this system, $\hat{H}_{0}(r)$ and $\hat{h}(r,t)$ are the $2N\times 2N$ matrix operators representing $H_{0}$ and $V_{C}$ [Eq. (2) and (3)] on the angular grid. The elements of the matrix $\hat{H}_{0}(r)$ are defined by
\begin{equation}
H_{0kj}^{ss^{'}}(r)=\lbrace-\frac{\hslash^{2}}{2\mu}\frac{\partial^{2}}{\partial\textit{r}^{2}}\delta_{kj}+(\lambda_{k}\lambda_{j})^{-1/2}
\sum_{\nu=(l,m)}^{N}(\varphi^{-1})_{k\nu}[V_{l}(r)+V_{l}^{s}(r)+\frac{\hslash^{2}l(l+1)}{2\mu\textit{r}^{2}}](\varphi^{-1})_{\nu\textit{j}}\rbrace\delta_{ss^{'}}\,,
\label{one}
\end{equation}

\noindent where $V_{l}(r)$ is the \textit{l}-dependent central potential between the neutron and the core and $V_{l}^{s}(r)$ is the spin dependent spin-orbit neutron-core interaction in the spin state $\vert\textit{s}>$~\cite{PRC:59(1999), PRC:68(2003), PRC50:1276(1994)}.
Notice that the time-dependent Coulomb operator $\hat{h}(r,t)$
\begin{equation}
h_{kj}^{ss^{'}}(r,t)=V_{C}(r,\Omega_{k},t)\delta_{ss^{'}}\delta_{kj}
\label{one}
\end{equation}

\noindent  defined in Eq. (3) is diagonal in this representation and does not require multipole expansion unlike  other approaches based on the numerical integration of the Schr\"odinger equation for the halo neutron~\cite{PRC50:1276(1994), NPA:581(1995)}.

The discrete-variable representation (9), (10) permits different kinds of modeling \textit{l}- and \textit{s}-dependent interactions $V_{l}(r)$ and $V_{l}^{s}(r)$ between the core and the halo neutron. It allows one to include resonance interactions in various partial \textit{l} and spin \textit{s} states in the $^{11}$Be Hamiltonian (9). We neglect here the spin-mixing term in the interaction between the core and the nucleon during the collision with the target following the conclusion made in~\cite{PRC:59(1999)}.  The computational scheme for integration of the system (8) is based on the splitting up method, as described in ~\cite{VSM:(1998), PRC:59(1999)}. The splitting up method gives a fast convergence with respect to the numbers of grid points $N$ (the number of basis functions in expansion (4)). It allows one to extend the boundaries of integration over radial and time variables to necessary values defined by the demand of accuracy.

For discretizing with respect to the radial variable \textit{r}, a sixth-order (seven point) finite-difference approximation on a quasiuniform grid has been used on the interval $r\in[0,r_{m}]$ with $r_{m}$=1200 fm.  The grid has been realized by mapping $r\rightarrow$\emph{x} the initial interval onto $x\in[0,1]$ by the formulas $r=r_{m}(e^{8x}-1)/((e^{8}-1))$~\cite{PLA:230(1997)}. The eigenvalue problem for the initial bound state and the final scattering states was calculated on the same radial grid.

\subsection{Model inputs}
\label{sec:helicity}

Our construction of the interaction $V(r)$ between the neutron and the $^{10}$Be core is based on the parametrization used in~\cite{PRC70:064(2004)} and represents the standard sum of the spherical Woods-Saxon potential $V_{l}(r)=V_{l}f(r)$, where $f(r)=1/(1+exp((r-R_{0})/a))$  with the radius $R_{0}$=2.585 fm and the diffuseness \textit{a}=0.6 fm, and the spin-orbit interaction

\begin{equation}
V_{l}^{s}(r)=V_{ls}\frac{1}{r}\frac{d}{dr}f(r)(\textit{\textbf{ls}})\,.
\label{one}
\end{equation}

\noindent The regular value $V_{ls}$=21 MeV $fm^{2}$ is used for the depth of the spin-orbit coupling potential (11) for a p-shell nucleus~\cite{PRC50:1276(1994)}. The Woods-Saxon potential with $V_{l}$=62.52 MeV (\textit{l}=0) and $V_{l}$=39.74 MeV (\textit{l}=1) reproduces the $1/2^{+}$ ground state of $^{11}$Be at -0.503 MeV, the $1/2^{-}$ excited state at -0.183 MeV and two resonant states $5/2^{+}$ and $3/2^{+}$ (see Table I) with the resonant parameters corresponding to theoretical~\cite{PAN:77(2014)} and experimental~\cite{NNDC} values.
To fix the position of the $3/2^{-}$ resonance close to the resonant value $E_{th}$=2.789 MeV from~\cite{PAN:77(2014)} we tuned the parameters of the Woods-Saxon potential ourselves (see Table II) .

\begin{table}[htp]
\caption{Theoretical and experimental parameters of resonant states of $^{11}$Be}
\begin{ruledtabular}
\begin{tabular}{ccccccc}

\multirow{2}{*}{} & \multicolumn{2}{c}{$5/2^{+}$} & \multicolumn{2}{c}{$3/2^{-}$} & \multicolumn{2}{c}{$3/2^{+}$} \\
 & E, MeV & $\Gamma$, keV & E, MeV & $\Gamma$, keV & E, MeV  & $\Gamma$, keV \\

\colrule
Theory~\cite{PAN:77(2014)} & 1.230 & 100  & 2.789 & 240 & 3.367 & 3 \\
 Experiment~\cite{NNDC} & 1.281 & 120  & 2.898 & 120 & 2.387 & 8 \\
\end{tabular}
\end{ruledtabular}
\end{table}


\begin{table}[htp]
\caption{Parameters of the potentials $V_{l}(r)$.}
\begin{ruledtabular}
\begin{tabular}{ccccccc}

& $ V_{l,even} (MeV) $ & $ V_{l,odd} (MeV) $ & $V_{ls} (MeV fm^{2}) $ & $ \textit{a} (fm)$ &
$ R_{0} (fm) $ & $ States $  \\
\colrule
Work~\cite{PRC70:064(2004)} & 62.52 & 39.74 & 21.0  & 0.60  & 2.585   &  $1/2^{\pm}, 5/2^{+}, 3/2^{+} $ \\
  Present work & & $6.80$ & 21.0 & $0.35$  & $2.500$  & $3/2^{-}$  \\
\end{tabular}
\end{ruledtabular}
\end{table}

\subsection{Breakup cross section}

The time evolution of the system is calculated according to the above scheme starting with the initial state $\Psi(\textbf{r},T_{in})=\phi_{2sj}(\textbf{r})$, where $\phi_{2sj}(\textbf{r})$ is the 2s - wave function of the projectile Hamiltonian $H_{0}(\textbf{r})$ (the ground state of $^{11}$Be)

\begin{equation}
H_{0}\phi_{ljm}(E,\textbf{r})=E\phi_{ljm}(E,\textbf{r})\,
\label{one}
\end{equation}

\noindent with $j=1/2$, where the projectile total momentum \textit{j} results from the coupling of the orbital momentum of the neutron \textit{l} with the spin of the neutron \textit{s}.

One can obtain the breakup component by eliminating the bound states from the calculated wave packet~\cite{PRC50:1276(1994), PRC:53(1996)}

\begin{equation}
|\Psi_{bu}(\textbf{r},t)\rangle=(1-\sum_{\nu\in\textit{bound}}|\phi_{\nu}(\textbf{r})\rangle\langle\phi_{\nu}(\textbf{r})|)|\Psi(\textbf{r},t)\rangle \,,
\label{one}
\end{equation}

\noindent where the sum runs over two bound states of $^{11}$Be calculated by integration of Eq.(12). Then the total breakup cross section can be calculated as a function of the relative energy \textit{E} between the emitted neutron and the core nucleus by the formulas

\begin{equation}
\frac{d\sigma_{bu}(E)}{dE}=\frac{4\mu\textit{k}}{\hslash^{2}}\int_{b_{min}}^{b_{max}}\sum_{lm}|\int\textit{j}_{l}(kr)Y_{lm}(\hat{r})\Psi_{bu}(\mathbf{r},T_{out})d\mathbf{r}|^{2}bdb \label{one}
\end{equation}

\noindent following the work ~\cite{PRC:59(1999)}. Here $\textit{j}_{l}(kr)$ is a regular spherical Bessel function representing the \textit{l}-wave component of the neutron wave function in the continuum spectrum $\textit{E}>$0 ($k=\sqrt{2\mu\textit{E}}/\hslash$) if the interaction between the core and the neutron is neglected.

Since one of the main objectives of this work is to study the influence of resonant states of $^{11}$Be on the reaction of its breakup, it becomes necessary to take into account the resonant nuclear interaction of the neutron with the core in the continuum spectrum. Therefore, we also use alternative formulas for the breakup cross section including neutron interaction with the core~\cite{PRC:68(2003), PRC:64(2001)}

\begin{equation}
\frac{d\sigma_{bu}(E)}{dE}=\frac{4\mu\textit{k}}{\hslash^{2}}\int_{b_{min}}^{b_{max}}\sum_{lm}|\int\phi_{ljm}(k,r)Y_{lm}(\hat{r})\Psi(\mathbf{r},T_{out})d\mathbf{r}|^{2}bdb\,.
\label{one}
\end{equation}

\noindent Here $\phi_{ljm}(k,r)$ is the radial part of the eigenfunction of the Hamiltonian $H_{0}(r)$ (12) in the continuum spectrum ($E=k^2\hbar^2/(2\mu)>0$), normalized to $\textit{j}_{l}(kr)$ as $kr\rightarrow\infty$ if $\textit{V(r)}$=0.

The initial (final) time $T_{in}$($T_{out}$) has to be sufficiently big $\vert\textit{T}_{in}\vert, T_{out}\rightarrow+\infty$, fixed from the demand for the time-dependent potential $V_{C}(\textbf{r},t)$ to be negligible at the beginning (end) of the  time-evolution. Following the investigation performed in Ref.~\cite{PRC:59(1999), PRC:68(2003)}, we fixed the time-interval as $T_{in}$= -20 $\hslash$/MeV and $T_{out}$=20 $\hslash$/MeV, which gives an accuracy of the order of one percent in calculating the breakup cross sections (14) and (15).

The boundary of integration over the impact parameter \textit{b} in formulae (14) and (15) were also chosen from the demand of accuracy to be of the order of one percent.
Table III illustrates that the demanded accuracy in computing the integral (14) is achieved as $b_{min}$= 12 fm and $b_{max}$=400 fm  for a few relative energies \textit{E}.
It should be noted that the inclusion of the region [0, $b_{min}$] makes sense if the nuclear interaction between the target and the projectile is taken into account~\cite{PRC:68(2003)}.

The quasiuniform radial grid with 2000 mesh points (generated by the step $\bigtriangleup\textit{x}=5\cdot10^{-3}$) and the edge at $r_{m}$=1200 fm gives the accuracy of integration of the order of about 1$\%$. The step of integration over the time variable $\bigtriangleup\textit{t}$=0.02  $\hslash$/MeV chosen in  ~\cite{PRC:59(1999)} keeps the same order of accuracy.


\begin{table}[htp]
\caption{Convergence of the breakup cross section $d\sigma(E,b_{max})/dE$ (in b/MeV) at 10 MeV/nucleon over the edge of integration $b_{max}$ (in fm) in Eq. (14) for different relative energies \textit{E} (in MeV). The calculations are performed for $N=81$ with including all three resonant states in the model.}
\begin{ruledtabular}
\begin{tabular}{cccccccccc}

 $b_{max}$ & E=0.1 & E=0.4  & E=0.8  &
E=1.2 & E=1.6 & E=2.0 & E=2.7 & E=3.0  & E=3.3  \\
\colrule
\textit{12} & 0.031 & 0.055 & 0.078 & 0.102 & 0.058 & 0.045 & 0.025 & 0.019 & 0.014 \\
\textit{20} & 0.350 & 0.696 & 0.842 & 0.767 & 0.413 & 0.273 & 0.121  & 0.085 & 0.060 \\
\textit{50} & 1.363 & 2.439 & 1.889 & 1.254 & 0.615 & 0.363 & 0.145 & 0.099 & 0.068 \\
\textit{100} & 1.839 & 2.809 & 1.966 & 1.269 & 0.618 & 0.364 & 0.145 & 0.099 & 0.068 \\
\textit{200} & 1.904 & 2.822 & 1.967 & 1.269 & 0.618 & 0.364 & 0.145 & 0.099 & 0.068 \\
\textit{400} & 1.907 & 2.826 & 1.968 & 1.269 & 0.618 & 0.364 & 0.145 & 0.099 & 0.068 \\
\end{tabular}
\end{ruledtabular}
\end{table}

The convergence of the method with respect to angular grid points $N\rightarrow\infty$ (number of basis functions (5) in expansion (4)) is discussed below, where it is investigated for the intermediate and lower beam energies.

\section{Results and discussion}
\subsection{Influence of resonant states on the breakup cross section of $^{11}$Be
at 69 and 72 MeV/nucleon.}

In the works ~\cite{PRC:59(1999)} and ~\cite{PRC:68(2003)}, the breakup reaction $^{11}$Be+$^{208}$Pb $\rightarrow$ $^{10}$Be+n+$^{208}$Pb was successfully investigated at 69 and 72 MeV/nucleon with the nonperturbative time-dependent approach we use here. However, the resonant states of $^{11}$Be were not included in these calculations.
Here, we overcome this drawback of the model: the resonant states $5/2^{+}$, $3/2^{-}$ and $3/2^{+}$ (see Table I) are taken into consideration. Note that these resonances were also observed experimentally by Fukuda et.al. in the breakup of $^{11}$Be on the light target ($^{12}$C) at 70 MeV/nucleon ~\cite{PRC70:054(2004)}.

\begin{figure}[htp]
\centerline{\includegraphics[width=9.8cm]{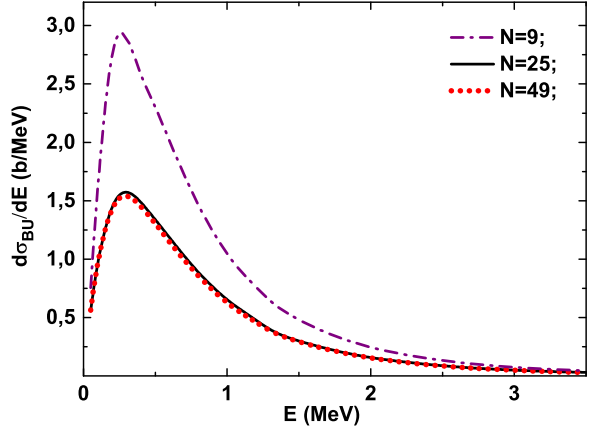}}
\caption{ The convergence of the breakup cross section $d\sigma_{bu}(E)/dE$ over the number $N$ of angular grid points calculated by Eq. (14) with including three resonant states between the neutron and the $^{10}$Be-core at 69 MeV/nucleon.}
\label{Fig:F2H}
\end{figure}
\begin{figure}[h!]
\centerline{\includegraphics[width=9.8cm]{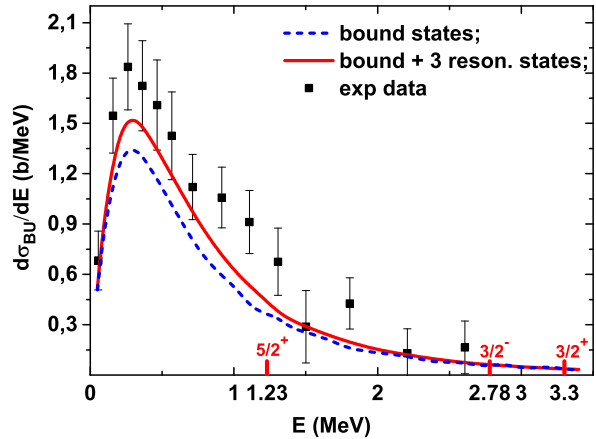}}
\caption{The breakup cross sections $d\sigma_{bu}(E)/dE$ calculated by Eq.(14) with including into interaction between the neutron and the $^{10}$Be-core only $1/2^{+}$ and $1/2^{-}$ bound states, and bound and resonant states ($5/2^{+}$, $3/2^{-}$, $3/2^{+}$) in comparison with experimental data at 72 MeV/nucleon~\cite{PLB:331(1994)}. The calculations are performed on the angular grid with $N=25$ grid points. Convolution of the calculation with the experimental resolution was not performed.}
\label{Fig:F2H}
\end{figure}

In Fig. 1, we illustrate the convergence of the computational scheme in calculating the cross section $d\sigma_{bu}(E)/dE$ for a beam energy 69 MeV/nucleon with respect to the number of angular grid points $N$ (the number of basis function (5) in expansion (4)). It is shown that the approach achieves the convergence at $N=25$.

\begin{table}[htp]
\caption{The contribution for different relative energies \textit{E} (in MeV) of the resonant states $5/2^{+}, 3/2^{-}$ and $3/2^{+}$ to the breakup cross section $d\sigma_{bu}(E)/dE$ (in b/MeV) at 69 MeV/nucleon. Here, 'b.s.' indicates the cross sections calculated with the interaction potential between the neutron and the $^{10}$Be-core including only two bound states of $^{11}$Be. }
\begin{ruledtabular}
\begin{tabular}{cccccc}
Energy &  \multicolumn{5}{c}{} \\
  & b.s. & b.s.+$5/2^{+}$ & b.s.+$5/2^{+}$+$3/2^{+}$ & b.s.+$5/2^{+}$+$3/2^{-}$ & b.s.+3 res. \\
\colrule
 E=0.1 & 0.936 & 0.936 & 1.007  & 1.011 & 1.011  \\
 E=0.3 & 1.420 & 1.421 & 1.595  & 1.606 & 1.606   \\
 E=0.8 & 0.704 & 0.708 & 0.865 & 0.875 & 0.875   \\
 E=1.0 & 0.555 & 0.562 & 0.639 & 0.647 & 0.648   \\
 E=1.2 & 0.383 & 0.378 & 0.473 & 0.479 & 0.479 \\
 E=2.0 & 0.138 & 0.137 & 0.155 & 0.156 & 0.156  \\
 E=2.7 & 0.056 & 0.056 & 0.070 & 0.069 & 0.069  \\
 E=3.0 & 0.043 & 0.044 & 0.051 & 0.050 & 0.050   \\
 E=3.3 & 0.036 & 0.036 & 0.037 & 0.037 & 0.037   \\
\end{tabular}
\end{ruledtabular}
\end{table}

In Fig. 2, we demonstrate that the inclusion of the resonant states in the interaction between the neutron and the $^{10}$Be-core gives a considerable contribution to the breakup cross sections. Thus, for a relative energy $E$= 1.2 MeV the 25\% increase of the cross section is observed, for $E$= 3.3 MeV the increment is about 3\%. Overall, it is shown that the  inclusion of the resonant states improves the agreement of the calculated breakup cross section with experimental data at 72 MeV/nucleon ~\cite{PLB:331(1994)}.

To clarify the contribution of the dominant resonance to the breakup cross section, we performed the computation at 69 MeV/nucleon, where most detailed and accurate experimental data are available~\cite{PRC70:054(2004)}. The partial contribution of each resonance is illustrated in Table IV. It is shown that the resonances $3/2^{-}$ and $3/2^{+}$ make a slightly larger contribution to the cross section $d\sigma_{bu}(E)/dE$ than $5/2^{+}$.

In Fig. 3, we demonstrate that including the resonant states between the neutron and the $^{10}$Be-core into the model improves the agreement of the calculated breakup cross sections with most accurate experimental data available at 69 MeV/nucleon~\cite{PRC70:054(2004)}. Here we also show that the effect of interaction of the neutron with the core at the final state of the breakup reaction (15) does not make a significant contribution to the cross section at intermediate energies around 70 MeV/nucleon. The influence of this effect on the breakup cross sections  for lower energies is discussed in the next subsection.

\begin{figure}[htp]
\centerline{\includegraphics[width=11cm]{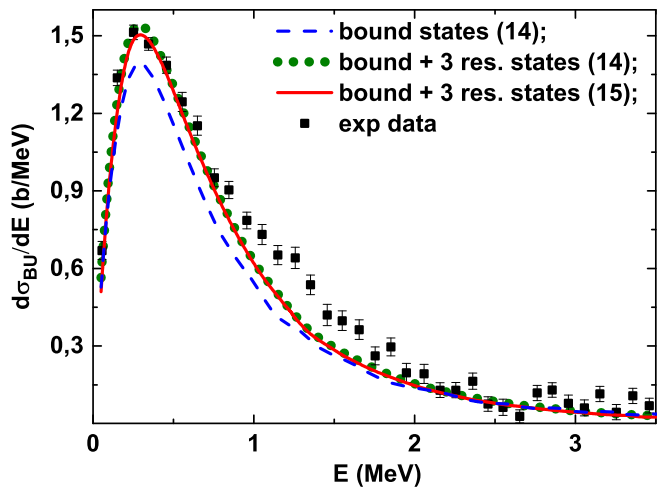}}
\caption{ Comparison of the calculated breakup cross sections with experimental data \cite{PRC70:054(2004)} at 69 MeV/nucleon. Calculations were performed by formulas (14) with only bound states in the interaction between the neutron and the $^{10}$Be-core (blue dashed curve) and with bound and  three resonant states ($5/2^{+}$, $3/2^{-}$, $3/2^{+}$) in the interaction (green dots). The case of including both bound and resonant states, as well as the interaction of the neutron with the core in the final state of the breakup reaction by (15) is also presented (red solid curve). Convolution of the calculation with the experimental resolution was not performed.}
\label{Fig:F2H}
\end{figure}

\subsection {Breakup cross section of $^{11}$Be at low beam energies}

Here we extend the approach for calculation of the breakup cross sections at low energy beams up to 5 MeV/nucleon and investigate the contribution of the $^{11}$Be resonance states in this region.

\begin{figure}[hbt]
\begin{minipage}[h]{0.5\linewidth}
\center{\includegraphics[width=1\linewidth]{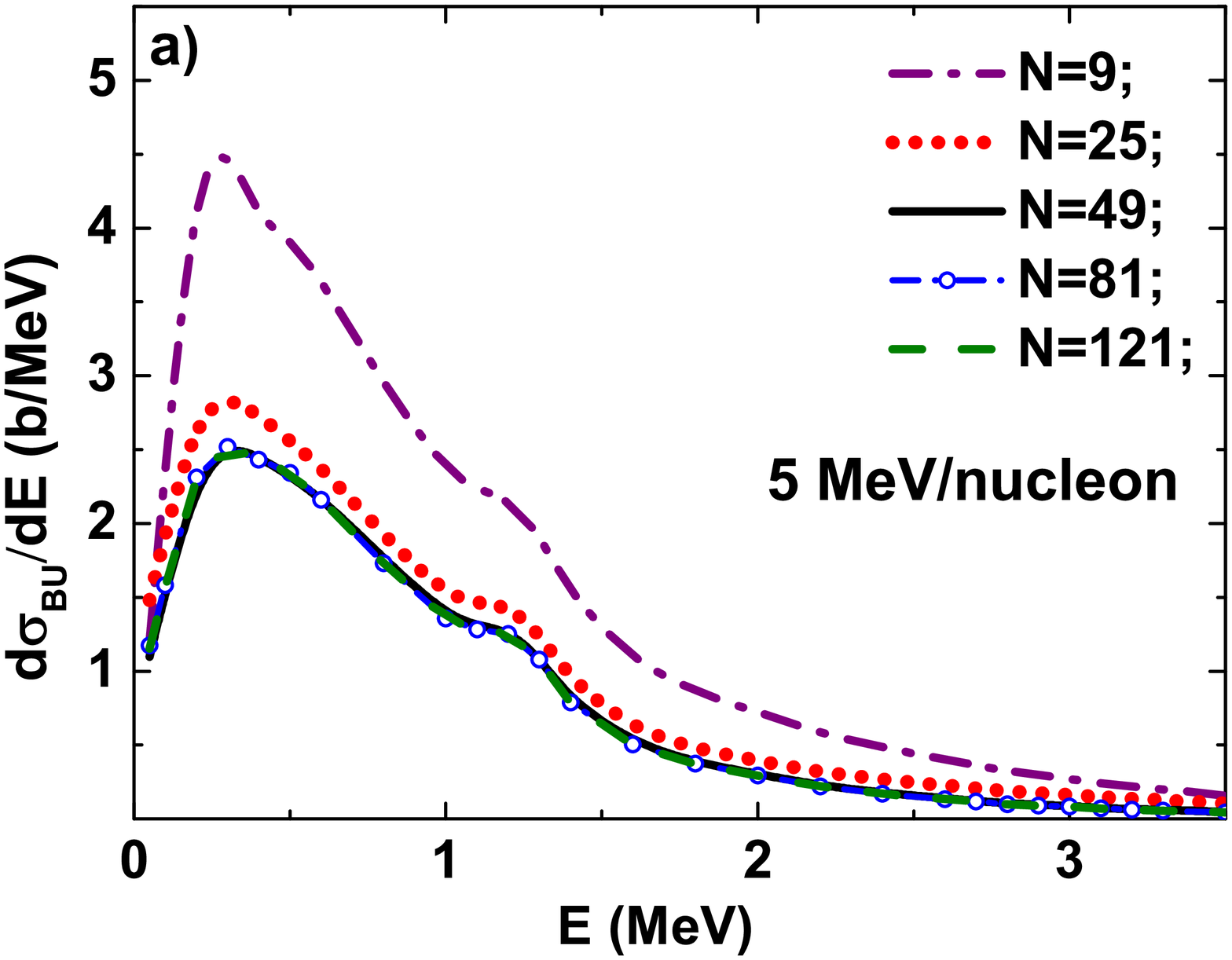}}
\end{minipage}
\hfill
\begin{minipage}[h]{0.49\linewidth}
\center{\includegraphics[width=1\linewidth]{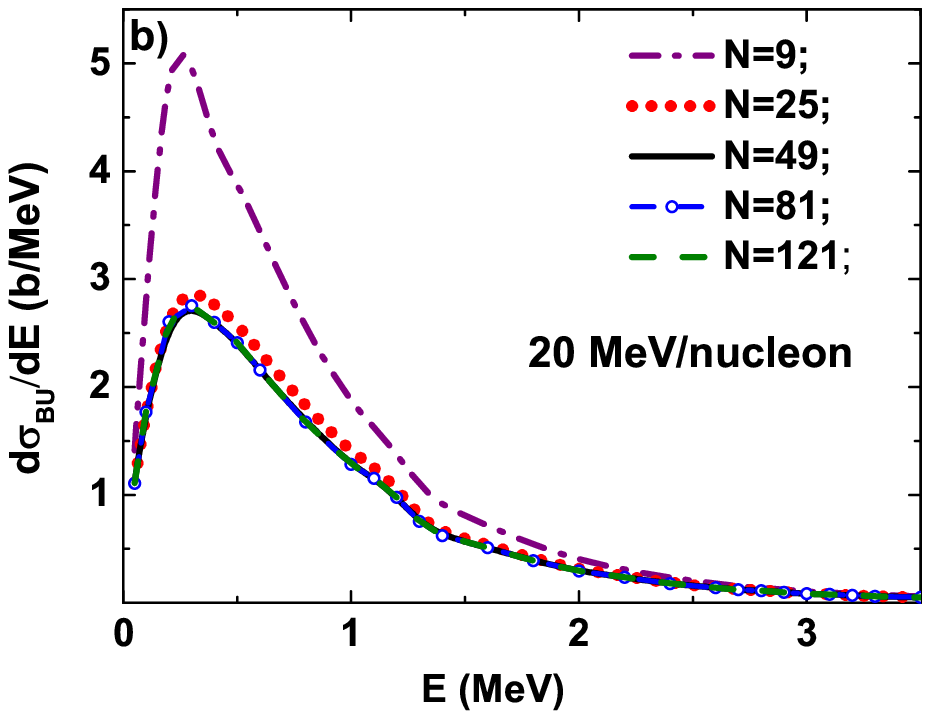}}
\end{minipage}
\caption{The convergence over the angular grid number $N$ of the calculated the breakup cross sections at 5 and 20 MeV/nucleon. The cross sections given in Fig. 4 were calculated  by formulas (14). The resonant states $5/2^{+}$, $3/2^{-}$ and $3/2^{+}$ were included into the computational scheme.}
\label{Fig:F2H}
\end{figure}

We first investigate the convergence of our computational scheme at low energies over the angular grid number $N$. For this, we calculate the breakup cross section $d\sigma_{bu}(E)/dE$ on different angular meshes at 5 and 20 MeV/nucleon. As it is shown at Fig. 4, for computing the breakup cross section at energies $\geq$ 20 MeV/nucleon with the demanded accuracy of the order of  one percent, it is sufficient to use $N=49$. For lower energies (up to 5 MeV/nucleon) the basis (the angular grid) should be extended to $N=81$ due to the slowing down of convergence.

\begin{figure}[h!]
\centerline{\includegraphics[width=10cm]{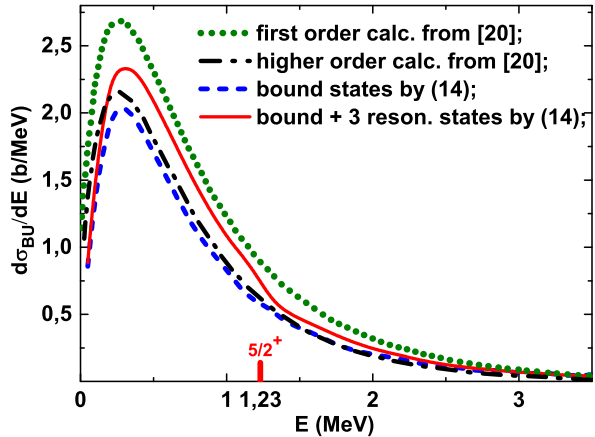}}
\caption{ The breakup cross section calculated by formulas (14) at 30 MeV/nucleon in comparison the CWBA calculations of~\cite{PRC:65(2002)}.}.
\label{Fig:F2H}
\end{figure}

In Fig. 5, we compare our results with the Coulomb wave Born approximation (CWBA) available for a beam energy of 30 MeV/nucleon~\cite{PRC:65(2002)}. Our calculations were performed with including only bound states (blue dash curve) and also with bound and three resonant states (red curve). They are compared with the CWBA calculations for the first order CWBA (green short dots) and the finite range CWBA (black dashed dots).
Figure 5 shows a significant deviation of the CWBA calculation from our result at 30 Mev/nucleon, which increases with decreasing  energy. This is consistent with the conclusion of the authors of ~\cite{PRC:65(2002)} about the difficulty of using the CWBA for lower energies. The resonant states of $^{11}$Be were not included in the CWBA calculations.

\begin{figure}[hbt]
\centerline{\includegraphics[width=11cm]{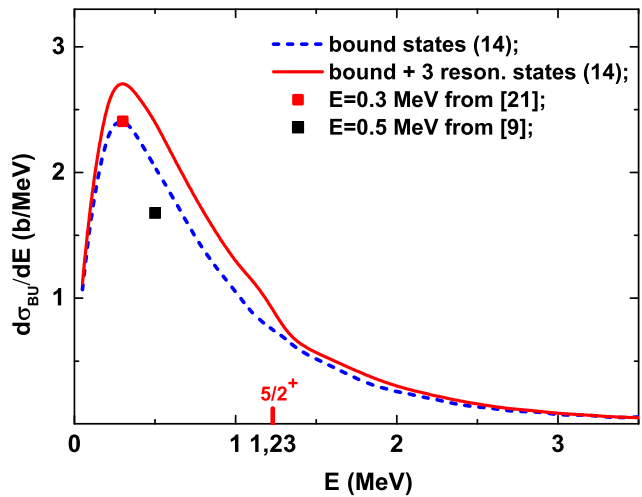}}
\caption{The breakup cross sections calculated with only bound states of $^{11}$Be in the computational scheme (blue dash curve) and with including three resonances of $^{11}$Be (red curve) at 20 MeV/nucleon by formulas (14) in comparison with calculations of the dynamical eikonal approaches at E = 0.3 ~\cite{PRC:73(2006)} and E = 0.5 MeV ~\cite{PRL:95(2005)}.}
\label{Fig:F2H}
\end{figure}

In Fig. 6 we compare our results with the breakup cross section calculated in~\cite{PRL:95(2005), PRC:73(2006)} within the dynamical eikonal approximations at fixed relative energies E= 0.3 and 0.5 MeV at 20 MeV/nucleon. In the dynamical eikonal calculations the resonant states of $^{11}$Be were not taken into account. Note that at an energy E= 0.3 MeV near the peak of the cross section, our calculation without including the resonant states of $^{11}$Be gives a cross section rather close to the dynamical eikonal approach ~\cite{PRC:73(2006)}.

\begin{figure}[hbt]
\begin{minipage}[h]{0.5\linewidth}
\center{\includegraphics[width=1\linewidth]{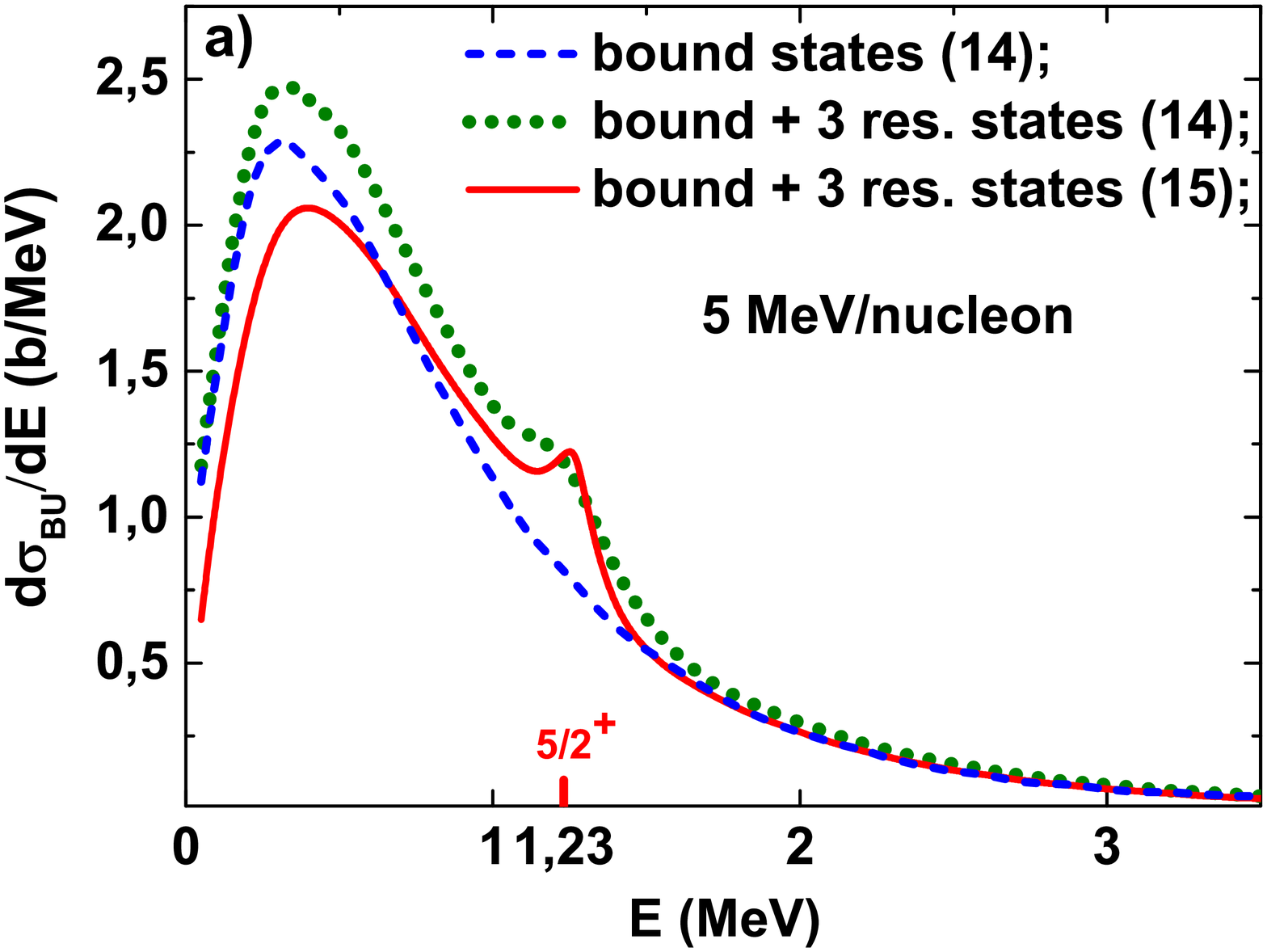}}
\end{minipage}
\hfill
\begin{minipage}[h]{0.49\linewidth}
\center{\includegraphics[width=1\linewidth]{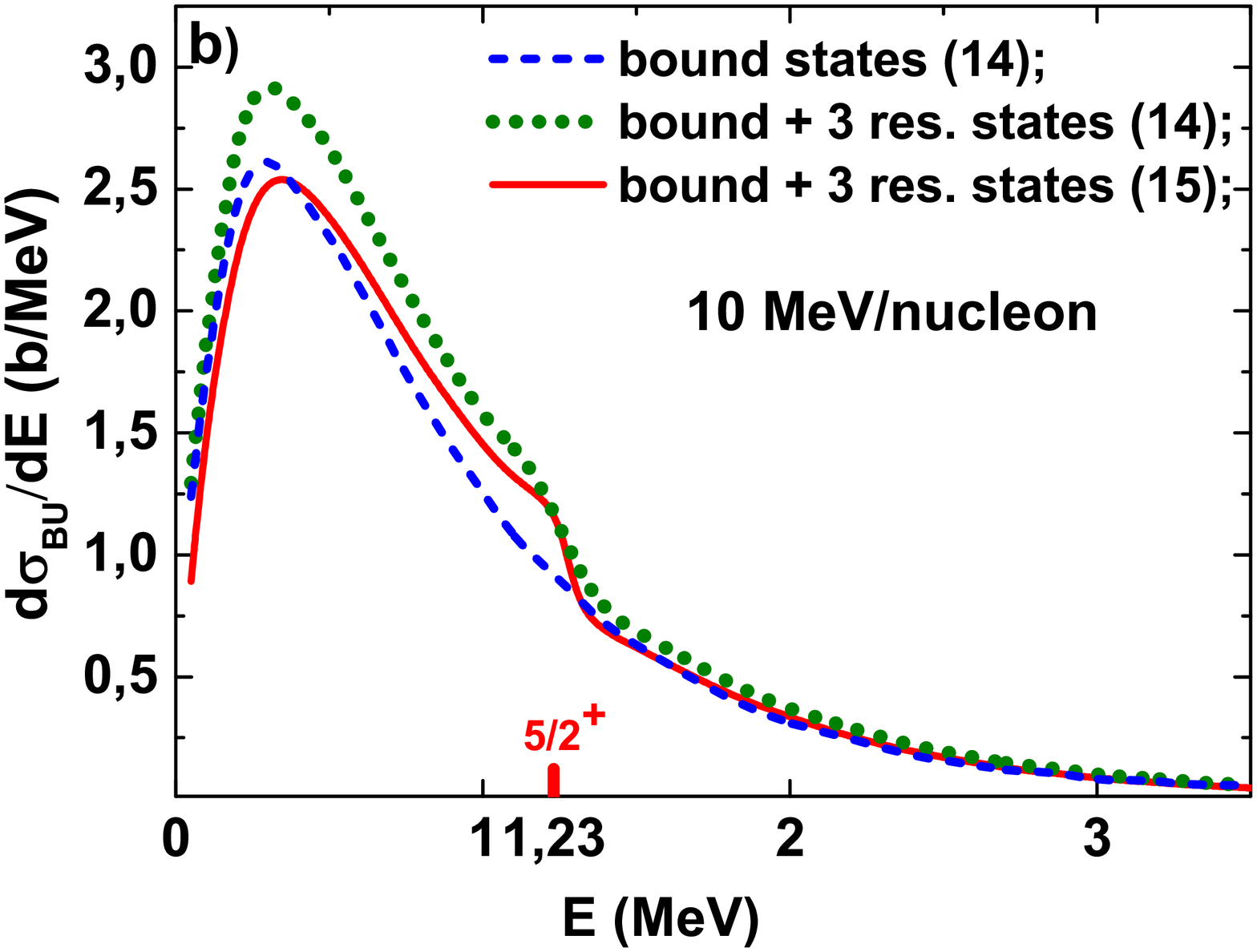}}
\end{minipage}
\caption{The contribution to the breakup cross section of the resonant states and the neutron interaction with the core in the continuum at beam energies of 5 and 10 MeV/nucleon. The calculations are performed at $N=81$.}
\label{Fig:F2H}
\end{figure}

Figure 7 demonstrates the contribution to the breakup cross sections of the resonant states $5/2^{+}$, $3/2^{-}$ and $3/2^{+}$ at beam energies of 5 (side a) and 10 MeV/nucleon (side b). The breakup cross sections were calculated by formulas (14), which approximated the continuum spectrum of the neutron by the Bessel functions (dashed and dotted curves). Calculations with formulas (15) taking into account the interaction of the neutron with the $^{10}$Be-core at the final state (red curve) are also presented here. It is shown that the inclusion of the neutron-core interaction in the final state of the breakup reaction considerably corrects the breakup cross sections, especially near the resonant energy 1.23 MeV of the  $5/2^{+}$ resonance.

\begin{figure}[h!]
\centerline{\includegraphics[width=11cm]{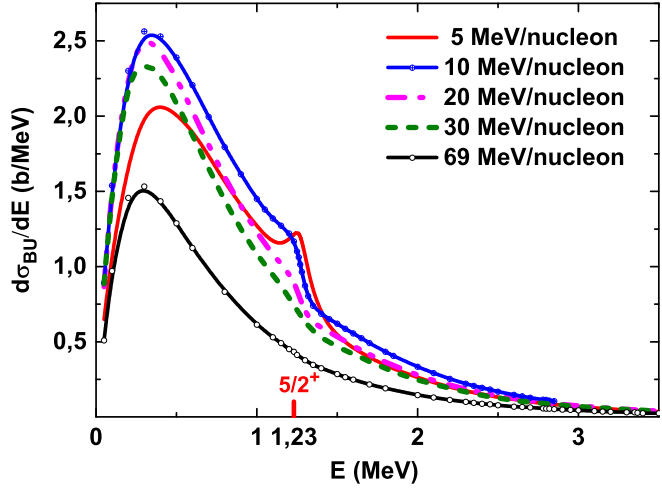}}
\caption{ The breakup cross section $d\sigma_{bu}(E)/dE$ calculated for different beam energies by formulas (15). In the calculation the bound and three resonant states were included in the neutron-core interaction.}
\label{Fig:F2H}
\end{figure}

Overall, the relative energy spectra of the fragments (neutron and core) were calculated for the Coulomb breakup of $^{11}$Be on the $^{208}$Pb target in the range 5 - 70 MeV/nucleon of beam energies.
In Fig. 8 we present the results of the calculations, which take into account the influence of the resonant states ($5/2^{+}$, $3/2^{-}$, $3/2^{+}$) and the effect of the neutron-core interaction in the final state (15) to the breakup cross section of the $^{11}$Be nucleus. The analysis performed demonstrates a strong dependence of the calculated cross sections on the beam energy and an increase in their sensitivity to low-lying resonance $5/2^+$ with decreasing energy. This also confirms the possibility of studying low-lying resonances in halo nuclei using their breakup reactions.

\subsection{Contribution to breakup of nuclear interaction between projectile and target}

The interaction $V_{C}(\textbf{r},t)$ of the target with the projectile (Eq.(3) in Subsection II.A) assumed to be purely Coulombic.	As it was shown in previous studies with time-dependent nonperturbative approach~\cite{PRC:68(2003)} for breakup reaction with a heavy target ($^{208}$Pb), the contribution of the nuclear part of the projectile-target interaction in the breakup cross sections is negligible around 70 MeV/nucleon. Recently, this effect was investigated at 69 MeV/nucleon using the CDCC method~\cite{PLB:811(2020)}. In this section we evaluate this effect at low beam energies following the approach of optical potential for the nuclear part $\Delta V_{N}(\textbf{r})=V_{cT}(r_{cT})+V_{nT}(r_{nT})$ between the target and projectile-nuclei interaction

\begin{equation}
V(\textbf{r},t)=V_{C}(\textbf{r},t)+\Delta V_{N}(\textbf{r})\,.
\label{one}
\end{equation}
Here $r_{cT}$ and $r_{nT}$ are the core-target $\textbf{r}_{cT}(t)=\textbf{R}(t)+m_{n}\textbf{r}/M$ and neutron-target $\textbf{r}_{nT}(t)=\textbf{R}(t)-m_{c}\textbf{r}/M$  relative variables and optical potentials $V_{cT}$ and $V_{nT}$ have the form:

\begin{equation}
V_{xT}(r_{xT})=-V_{x}f(r_{xT}, R_{R}, a_{R})-iW_{x}f(r_{xT}, R_{I}, a_{I})\,
\label{one}
\end{equation}
with Woods-Saxon form factors $f(r_{xT},R_{R},a_{R}) = 1/(1+exp(r_{xT}-R)/a)$, where $x$ stands for either core or neutron. We use here the parameters of the optical potentials (17) from the paper~\cite{PRC:68(2003)} which are given in Table V.


\begin{table}[htp]
\caption{Parameters of the core-target and neutron-target optical potentials.}
\begin{ruledtabular}
\begin{tabular}{ccccccc}

$c$ or $n$ & $ V_{x} (MeV) $ & $ W_{x} (MeV) $ & $R_{R} (fm) $ & $ R_{I} (fm)$ &
$ a_{R} (fm) $ & $a_{I} (fm) $  \\
\colrule
$^{10}$Be & 70.0 & 58.9 & 7.43  & 7.19 & 1.04 & 1.00 \\
$n$ & 28.18 & 14.28 & 6.93 & 7.47 & 0.75 & 0.58 \\
\end{tabular}
\end{ruledtabular}
\end{table}

At previous sections the  nuclear interaction effects were simulated by a cutoff $b_{min}$= 12 fm of the impact parameter at the Eqs.(14,15). The including of nuclear interaction $\Delta V_{N}(\textbf{r})$ between the projectile and the target afford to make a calculations with smaller impact parameters starting from $b_{min}$= 5 fm. The breakup cross sections calculated with Coulomb and additional nuclear part of interaction are compared in Fig. 9 for lower beam energies (5 and 30 MeV/nucleon). It is shown that the cutoff Coulomb approximation (3) underestimates the breakup cross section including the nuclear interaction between the projectile and the target (16). In the calculation, as well as in the previous sections, two bound states (ground $1/2^{+}$ and first excited $1/2^{-}$ states) and three resonance states ($5/2^{+}$, $3/2^{-}$ and $3/2^{+}$) of $^{11}$Be were taken into account. The calculation were performed on the angular grid giving convergent results with N=121 grid points.

\begin{figure}[hbt]
\begin{minipage}[h]{0.5\linewidth}
\center{\includegraphics[width=1\linewidth]{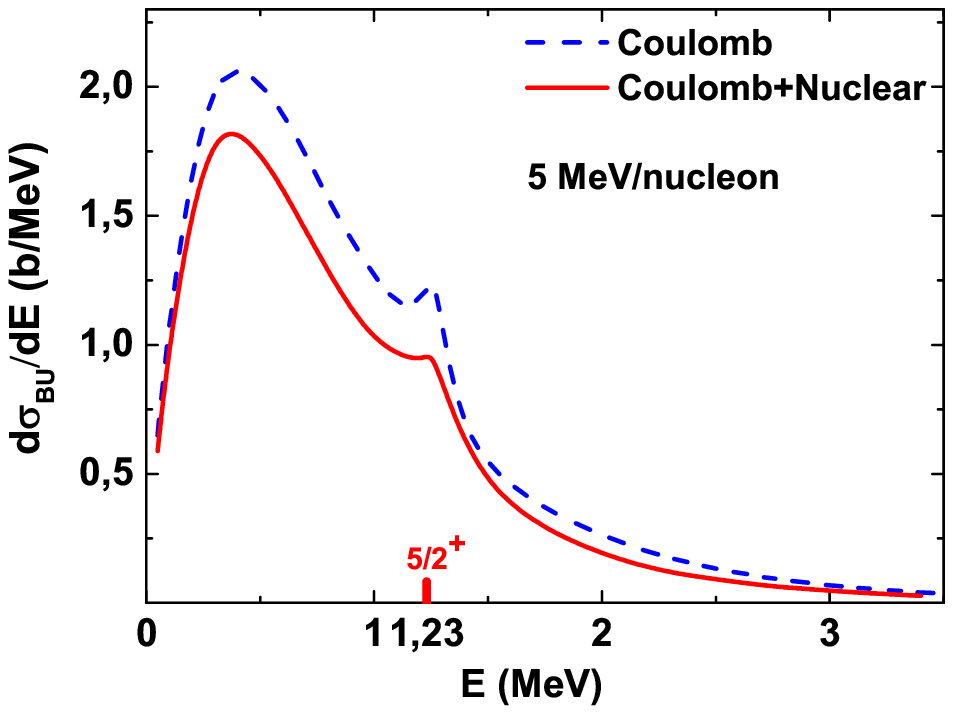}}
\end{minipage}
\hfill
\begin{minipage}[h]{0.49\linewidth}
\center{\includegraphics[width=1\linewidth]{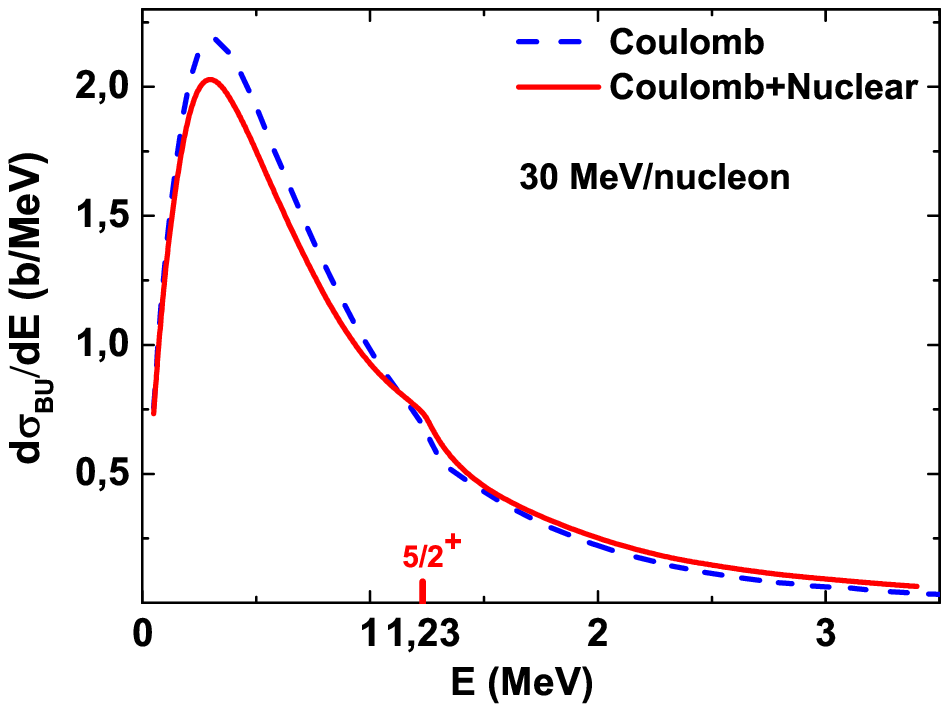}}
\end{minipage}
\caption{Breakup cross sections calculated with pure Coulombic projectile-target interaction (3) (dashed curves) and with the interaction (16) including nuclear effect (full curves) for beam energies 5 and 30 MeV/nucleon.}
\label{Fig:F2H}
\end{figure}

In conclusion, we see that with a decreasing of a beam energy, the influence of the nuclear effect in the projectile-target interaction becomes more significant for the breakup cross section. The contribution from the resonance states remains noticeable when the nuclear interaction between the target and the projectile is included, and the peak due to the $5/2^{+}$ resonance is still clearly visible at low beam energies. Note that a noticeable manifestation of nuclear effects in the differential cross section for the breakup of the $^{11}$Be on $^{208}$Pb at 12.7 MeV/nucleon was also pointed out in the recent work~\cite{PLB:811-2(2020)}, where the experimental data were analyzed with the CDCC method.

\subsection{Excitation of $^{11}$Be in collision with $^{208}$Pb}

We have also calculated the transition probability $P_{lj}$ during the collision of $^{11}$Be  with $^{208}$Pb to the excited state $1/2^{-}$ with the energy $E_{l=1,j=1/2}$ = -0.183 MeV

\begin{equation}
P_{lj}(b)=\frac{1}{2}\sum_{m}|\langle\phi_{ljm}(E_{lj},r)|\Psi(\mathbf{r},T_{out}\rangle|^{2}\,
\label{one}
\end{equation}
and the corresponding inelastic cross section

\begin{equation}
\sigma_{exc}(E)=2\pi\int_{0}^{\infty}P_{lj}(b)bdb\,.
\label{one}
\end{equation}

First, in order to evaluate the contribution of the $E1$ transition in the excitation cross sections (19), we perform calculations using only the electric dipole term  in the time-dependent potential (3). In other words, the time-dependent Coulomb potential is replaced by the first term of its multipole expansion:

\begin{equation}
V_{C}^{E1}(\textbf{r},t)=-Z_{C}Z_{T}e^{2}\dfrac{m_{n}}{M}\dfrac{(\textbf{r}\cdot\textbf{R}(t))}{R(t)^{3}}\,.
\label{one}
\end{equation}
The transition probabilities (18) and cross sections (19) were also calculated with pure Coulomb projectile-target interaction (3) and with included nuclear effects between the target and the projectile (16). These results are presented in Fig. 10 for beam energy 69 MeV/nucleon. They demonstrate a pure Coulomb mechanism of excitation of the $1/2^{-}$ state of $^{11}$Be with the overhelming dominance of the $E1$ transition for impact parameters $b>15$ fm. With a decrease of $b$, the excitation probability of the $1/2^{-}$ state sharply decreases due to the influence of the nuclear interaction between the target and the projectile.

\begin{figure}[htp]
\centerline{\includegraphics[width=11cm]{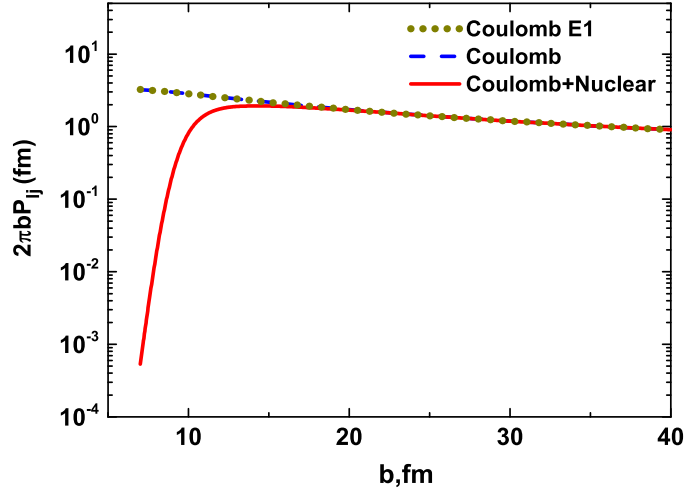}}
\caption{The transition probabilities $P_{lj}(b)$ (multiplied by 2$\pi$b) in the collision of $^{11}$Be with $^{208}$Pb to the $1/2^{-}$ state, calculated with Coulomb plus optical potential (full line) and pure Coulomb excitation (dashed lines) in comparison with $E1$ dipole excitation (dotted line) at 69 MeV/nucleon.}
\label{Fig:F2H}
\end{figure}

\begin{figure}[htp]
\centerline{\includegraphics[width=11cm]{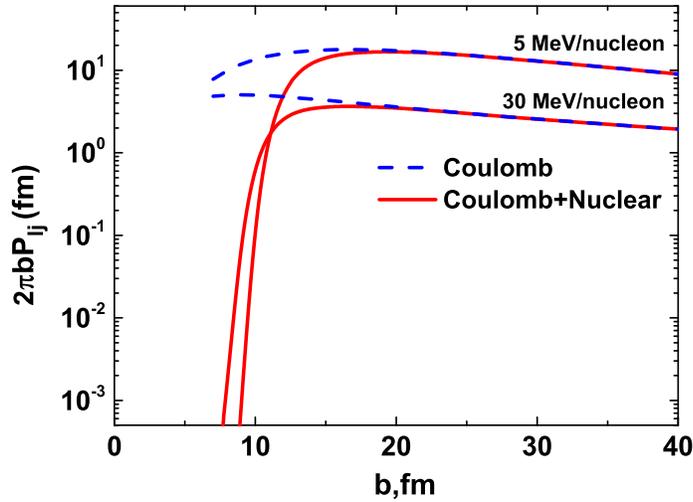}}
\caption{ Excitation probabilities $P_{lj}(b)$ (multiplied by 2$\pi$b) of $^{11}$Be in the collision with $^{208}$Pb to the $1/2^{-}$ state, calculated with Coulomb plus optical potential (full line) and pure Coulomb excitation (dashed lines) at 5 and 30 MeV/nucleon.}
\label{Fig:F2H}
\end{figure}

The excitation probabilities $P_{lj}(b)$ (multiplied by 2$\pi$b) calculated at low colliding energies are presented in Fig. 11. The probabilities calculated with including  nuclear effects are smaller than those with a pure Coulomb interaction between the target and projectile. We see the lower the projectile energy, the more influence has a nuclear effect.
For all considered projectile energies starting from $b=15-20$ fm the calculated excitation probabilities are determined by the Coulomb interaction between the projectile and the target. We may conclude that the choice of minimal impact parameter $b_{min}$= 12 fm as a cutoff approximation in the previous calculations for 72 MeV/nucleon at ~\cite{PRC:68(2003)} was realistic, since the contribution of nuclear effects completely reduces the transition probability below $b=12$ fm.

The calculated excitation cross section of the $1/2^{-}$ state of $^{11}$Be (19) are given in Fig. 12 as a function of the projectile energy per nucleon for pure Coulomb and Coulomb plus nuclear induced excitations. A monotonic increase of the inelastic cross section (19) with decreasing projectile energy is observed. Also, it is noted that the difference between the results with pure Coulomb projectile-target interaction and with adding optical potentials increases for lower beam energies.

\begin{figure}[htp]
\centerline{\includegraphics[width=11cm]{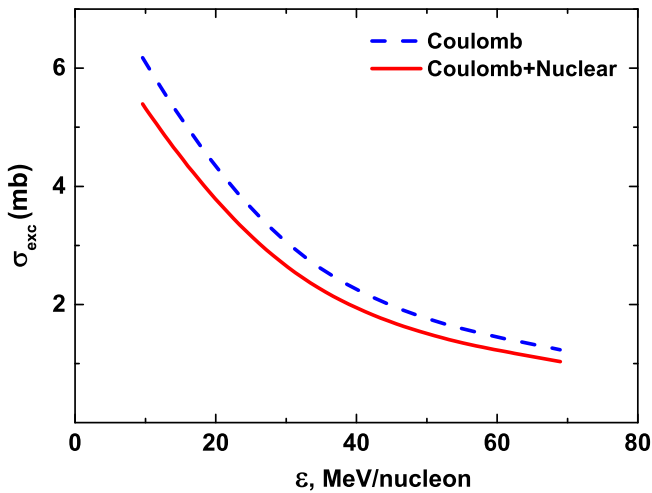}}
\caption{Inelastic cross section to the  $1/2^{-}$ excited state of $^{11}$Be as a function of the projectile energy per nucleon $\varepsilon$ (in MeV/nucleon) in comparison with pure Coulomb (dashed line) and with including nuclear interactions (full line) between the projectile and the target at the energy range 5 - 72 MeV/nucleon.}
\label{Fig:F2H}
\end{figure}

\section{Conclusion}
\label{sec:summary}

In this paper, the Coulomb breakup of one-neutron halo nuclei $^{11}$Be on the heavy target $^{208}$Pb has been studied within the nonperturbative time-dependent approach. The used three-dimensional mesh approximation (discrete-variable representation for the angular part and high order finite-differences for the radial part) has the advantage that any local interaction is diagonal in such a representation. As a consequence, it allows avoiding the use of the multipole expansion of the time-dependent Coulomb interaction between the projectile and the target and the straight-line approximation for the projectile trajectory ~\cite{PRA:69(2004)}. The splitting up procedure suggested in ~\cite{VSM:(1998)} and ~\cite{PRC:59(1999)} permitted us to include correctly the resonances in different partial and spin states of $^{11}$Be. Overall, this numerical technique allows an accurate and straightforward modelling of the nuclear interaction between the nucleon and the $^{10}$Be-core in a wide range of beam energies.

The relative energy spectra of the fragments (neutron and core) were calculated for the Coulomb breakup of $^{11}$Be on the $^{208}$Pb target in the range of beam energies 5 - 70 MeV/nucleon. In these calculations the influence of the resonant states $5/2^{+}$, $3/2^{-}$ and $3/2^{+}$ of $^{11}$Be on the breakup cross section was taken into account. These results have been compared with experimental data~\cite{PLB:331(1994), PRC70:054(2004)} available at 69 and 72 MeV/nucleon and with existing alternative calculations of other authors~\cite{PRC:65(2002), PRL:95(2005), PRC:73(2006)} at 20 and 30 MeV/nucleon.  In our calculations performed for the incident beam energies at 5 $–$ 30 MeV/nucleon region, the contribution of the $5/2^{+}$ resonance state of $^{11}$Be to the breakup cross sections is clearly visible, while at energies of 69 and 72 MeV/nucleon, the resonant states $3/2^{-}$ and $3/2^{+}$ give  the main contribution to breakup cross sections of $^{11}$Be. At 69 MeV/nucleon the similar effect was observed in \cite{PRC:89(2014)} in the differential cross sections: the main contribution of the resonance $3/2^{-}$. We have shown that the inclusion of the resonant states of $^{11}$Be into the computational scheme leads to a significant improvement of the theoretical model, which gives a better agreement of the model  description of the experimental data on the breakup cross sections ~\cite{PLB:331(1994), PRC70:054(2004)}. Also the contribution of nuclear interaction between the target and projectile to the breakup cross sections at low beam energies (5 - 30 MeV/nucleon) was evaluated. Our approach can potentially be used in further investigation of breakup reactions at low energies.

In addition, the developed theoretical approach could easily deal with nuclear effects between the target and the projectile, which are supposed to be important in the breakup of halo nuclei on light targets. Another interesting, but more difficult application of the developed theoretical model may be breakup reactions of two-neutron halo nuclei.

\section{Acknowledgments}

We acknowledge valuable discussion with Prof. S.N. Ershov. The work was supported by Grant No. 075-10-2020-117 of the Ministry of Education and Science of the Russian Federation.

\end{document}